\documentclass[%
reprint,
superscriptaddress,
amsmath,amssymb,
aps,
prb,
floatfix,
]{revtex4-2}

\usepackage{graphicx}
\usepackage{dcolumn}
\usepackage{bm}
\usepackage{hyperref}
\usepackage[mathlines]{lineno}
\usepackage{amsmath,bm}
\usepackage{times}
\usepackage{fouriernc}
\usepackage{xcolor}

\begin{document}

\preprint{APS/123-QED}

\title{Highly anisotropic magnetism in the vanadium-based kagome metal TbV$_6$Sn$_6$}

\author{Ganesh Pokharel}
\email{ganeshpokharel@ucsb.edu}
\affiliation{Materials Department, University of California Santa Barbara, Santa Barbara, California 93106, USA}
\author{Brenden Ortiz}
\affiliation{Materials Department, University of California Santa Barbara, Santa Barbara, California 93106, USA}
\author{Juan Chamorro}
\affiliation{Materials Department, University of California Santa Barbara, Santa Barbara, California 93106, USA}
\author{Paul Sarte}
\affiliation{Materials Department, University of California Santa Barbara, Santa Barbara, California 93106, USA}
\author{Linus Kautzsch}
\affiliation{Materials Department, University of California Santa Barbara, Santa Barbara, California 93106, USA}
\author{Guang Wu}
\affiliation{Department of Chemistry and Biochemistry, University of California Santa Barbara, Santa Barbara, California 93106, USA}
\author{Jacob Ruff}
\affiliation{CHESS, Cornell University, Ithaca, New York 14853, USA}
\author{Stephen D. Wilson}
\email{stephendwilson@ucsb.edu}
\affiliation{Materials Department, University of California Santa Barbara, Santa Barbara, California 93106, USA}

\date{\today}

\begin{abstract}
$R$V$_6$Sn$_6$ ($R$=rare earth) compounds are appealing materials platforms for exploring the interplay between $R$-site magnetism and nontrivial band topology associated with the nonmagnetic vanadium-based kagome network. Here we present the synthesis and characterization of the kagome metal TbV$_6$Sn$_6$ via single crystal X-ray diffraction, magnetization, transport, and heat capacity measurements. Magnetization measurements reveal strong, uniaxial magnetic anisotropy rooted in the alignment of Tb$^{3+}$ moments in the interplane direction below 4.3(2) K. TbV$_6$Sn$_6$ exhibits multiband transport behavior with high mobilities of charge carriers, and our measurements suggest TbV$_6$Sn$_6$ is a promising candidate for hosting Chern gaps driven via the interplay between Tb-site magnetic order and the band topology of the V-site kagome network.

\end{abstract}

\pacs{Valid PACS appear here}
\maketitle

\section{Introduction}
The exploration of correlated topological electronic states born from the interplay between lattice geometries/nontrivial band topologies and electron-electron interactions is an active frontier of condensed matter physics research \cite{Yin2018,Yangeabb6003,Noam_2020,Liu_2019,PhysRevLett.126.246602,Pokharel_2021,Tsai2020,Jiang2021,Zhao2021, ghimire_2020}. Transition-metal-based kagome metals have attracted particular attention due to their natural potential to host topologically nontrivial electronic states and a wide variety of electronic correlations \cite{review}. Predictions include quantum anomalous Hall states for dissipationless spintronics and topologically nontrivial superconducting phases sought for fault-tolerant quantum computing.  The observation of unusually large anomalous Hall responses \cite{Yangeabb6003, Kida_2011, liu_sun_2018}, complex itinerant magnetism \cite{Fenner_2009, Rebacca_2021}, charge density waves \cite{Jiang2021, https://doi.org/10.48550/arxiv.2203.11467, arachchige2022charge, Zhao2021}, and superconductivity \cite{ortiz2020superconductivity, ortizCsV3Sb5, 2021Rb} \emph{etc.} within kagome metals seemingly validates their promise as hosts of a rich frontier of unconventional ground states and phase behaviors. 

In particular, kagome metals that crystallize in the MgFe$_6$Ge$_6$ structural prototype, or 166-kagome metals, have recently been proposed as a chemically flexible route for exploring the interplay between topology, correlations and magnetism \cite{ghimire_2020,PhysRevLett.126.246602, https://doi.org/10.48550/arxiv.2203.09447, https://doi.org/10.48550/arxiv.2106.15494,Li_2021_dirac,PhysRevB.103.014416,Chern_Yin2020_TbV6Sn6,Xu2022,Pokharel_2021, Pokharel_2021_2}. In Mn-based 166-kagome lattices of the form $R$Mn$_6$Sn$_6$ ($R$= rare earth), the combination of magnetic symmetry breaking with nontrivial band topology stabilizes a variety of responses including spin polarized Dirac cones \cite{Li_2021_dirac}, large anomalous Hall effects \cite{Chen_Dong_2021,Asaba_2020}, and spin-orbit-driven Chern gap formation \cite{Chern_Yin2020_TbV6Sn6}. Notably, the inclusion of spin–orbit coupling and out-of-plane ferromagnetism in these materials is reported to generate dissipation-less electronic edge states within the gapped Dirac cones of TbMn$_6$Sn$_6$ \cite{Chern_Yin2020_TbV6Sn6, https://doi.org/10.48550/arxiv.2203.09447, https://doi.org/10.48550/arxiv.2106.15494}, and the dominant exchange field of the Mn-based kagome network is known to form a variety of magnetic structures \cite{CLATTERBUCK199978,MALAMAN1999519}. 

An appealing complement to these studies is to explore 166-kagome metals where the magnetic interactions/anisotropies can be tuned independent of the kagome lattice, and to, in turn, control the interplay between magnetic order and the kagome-derived band structure. Vanadium-based kagome metals of the form $R$V$_6$Sn$_6$ are recently identified compounds with a nonmagnetic V-based kagome lattice that allow for this control \cite{Pokharel_2021, Pokharel_2021_2, Japanese_166}. The presence of topological surface states, Dirac cones, and van Hove singularities near $E_F$ in these compounds support the idea of interfacing magnetism with the topological electronic features of the nonmagnetic kagome sublattice. While $R$=Gd manifests a noncollinear, relatively isotropic magnetic state \cite{Pokharel_2021}, exploring other variants with differing anisotropies, such as the typically uniaxial Tb ion, remains unexplored. 

In this work, we report the synthesis of single crystals of TbV$_6$Sn$_6$ and study their crystal structure and electronic properties using synchrotron X-ray diffraction, magnetic susceptibility, magnetotransport, and heat capacity measurements. X-ray diffraction data reveal a $P6/mmm$ hexagonal symmetry maintained to 10 K, and magnetization measurements reveal ferromagnetic order below $T_C=$4.3(2) K with a strong easy-axis alignment of Tb moments along the crystallographic c-axis.  Magnetotransport data show high mobility and multiband behavior at low temperature, and our results motivate TbV$_6$Sn$_6$ as a host of tunable Chern gaps due to the interplay between Tb-site magnetism, spin-orbit coupling, and the Dirac electronic spectrum of the V-site kagome nets.

\begin{figure*} 
\centering
\includegraphics[width=2\columnwidth]{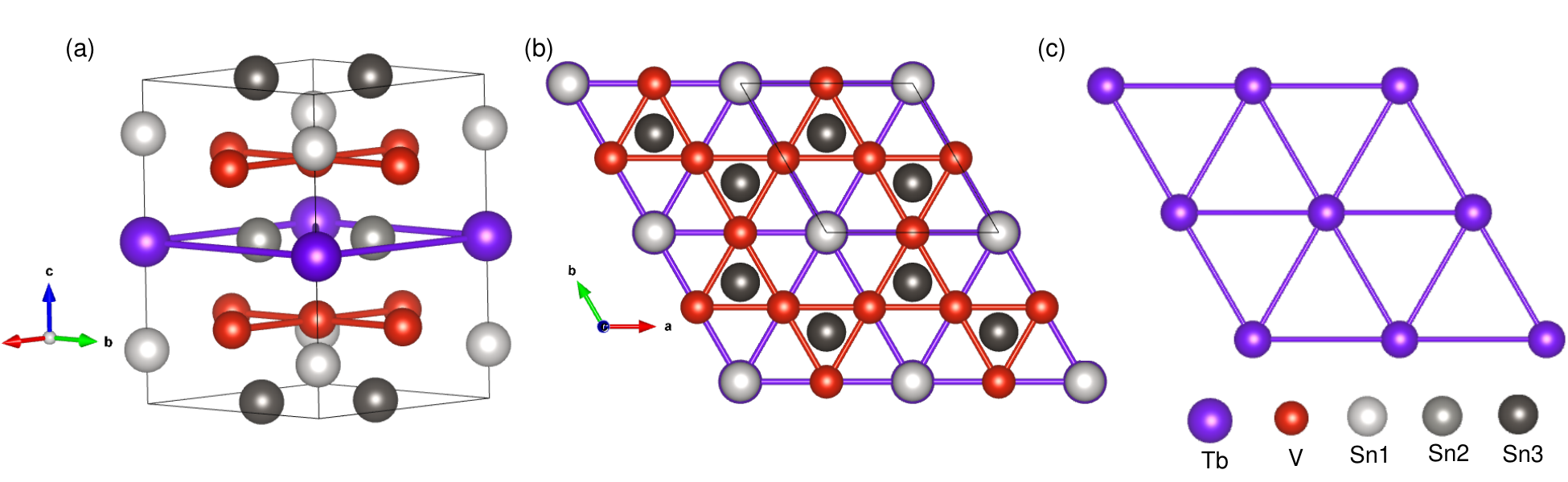}
      \caption{Crystal structure of TbV$_6$Sn$_6$. (a) A unit cell formed by the  unique arrangements of Tb, V and Sn atoms. Sn atoms
occupying the three different crystallographic sites are depicted by light black, medium black and dark black color spheres. (b) Top view of crystal structure, viewed along c-axis, showing the kagome layer of V-atoms and projected Sn1 and Sn3 sites. (c) Triangular lattice of Tb- ions interwoven between the kagome planes of V-ions. 
      }
\label{Fig_1}
\end{figure*}

\section{Experimental Details}

Single crystals of TbV$_6$Sn$_6$ were synthesized via a conventional flux-based growth technique. Tb (pieces, 99.9\%), Sn (shot, 99.99\%), V (pieces, 99.7\%) were loaded inside a ``Canfield-crucible" with a molar ratio of 1:6:20 and then sealed in a quartz tube filled with 1/3 atm of Ar. Canfield crucible sets consist of two cylindrical crucibles separated by a frit-disc, which are effective to separate the liquid from solid phase at the time of centrifuge. 1/3 Ar atm inside the tube is helpful to suppress volatility within the tube. The sealed tube was heated at 1125$^{\circ}$C for 12 hours and then slowly cooled down to 780 $^{\circ}$C at a rate of 2$^{\circ}$C/h. Plate-like single crystals were separated from the molten flux via centrifuging at 780 $^{\circ}$C. Crystals grown via this technique are usually a few millimeters in length and around 200 $\mu m$ in thickness. The surfaces of crystals were often contaminated with Sn and were subsequently washed with dilute HCl to remove the surface contamination.  YV$_6$Sn$_6$ crystals were synthesized in a similar manner \cite{Pokharel_2021}.

Laboratory X-ray diffraction measurements were performed on a Kappa Apex II single-crystal diffractometer with a fast charge-coupled device (CCD) detector and a Mo source. To obtain the structural solution, the laboratory X-ray diffraction pattern was refined using SHELX software package \cite{Shelx_2015}. Additional, temperature-dependent synchrotron X-ray diffraction were collected at ID4B (QM2) beamline, CHESS. In ID4B measurements, temperature was controlled by a stream of cold helium gas flowing across the single-crystal sample. An incident X-ray of energy 26 keV (wavelength  $\sim$ 0.67 $\AA$) was selected using a double-bounce diamond monochromator. Bragg reflections were collected in transmission mode, and the sample was rotated with full 360-degree patterns, sliced into 0.1-degree frames, and the data were then analyzed in the APEX3 software package. Final refinement of the crystal structure was done using the integrated SHELX and APEX3 software package. To further verify phase purity over a larger volume, powder X-ray diffraction measurements were carried out on a Panalytical Empyrean powder diffractometer using crushed single crystals. 

Magnetization measurements as a function of temperature and magnetic field were carried out using a Quantum Design Magnetic Properties Measurement Systems (MPMS-3). Prior to measurement, the plate-like crystals were polished gently on the top and bottom surfaces to remove any surface contamination. Single crystals of mass $\approx$ 1 mg were attached to quartz paddles using GE-Varnish. Magnetization data were collected in the temperature range of 1.8-300 K with the magnetic field applied parallel to and perpendicular to the crystal surface.  

Electrical transport measurements were carried out using the electrical transport option (ETO) of the Quantum Design Dynacool Physical Properties Measurement System (PPMS). A standard four-probe, Van der Pauw measurement method was used for the measurements of transverse magnetoresistance and Hall resistances, and the magnetic field was applied perpendicular to the direction of the current flow. Heat capacity measurements of a 1.3 mg single crystal were carried out using a Dynacool PPMS, and to account the phonon contributions to the heat capacity of TbV$_6$Sn$_6$, a YV$_6$Sn$_6$ 3.68 mg crystal was utilized.  The heat capacity of YV$_6$Sn$_6$ (Y166) was scaled as a phonon lattice standard for TbV$_6$Sn$_6$ using the relation $C_{phonon}=C_{Y166}\left(\frac{m_Y+6m_V+6m_{Sn}}{m_{Tb}+6m_V+6m_{Sn}}\right)^{3/2}$ with a scaled temperature $T_{nonmag}=T_{Y166}\left(\frac{m_Y+6m_V+6m_{Sn}}{m_{Tb}+6m_V+6m_{Sn}}\right)^{1/2}$ \cite{plumb2016antiferromagnetic}.

\begin{figure} 
	\centering
	\includegraphics[width=1\columnwidth]{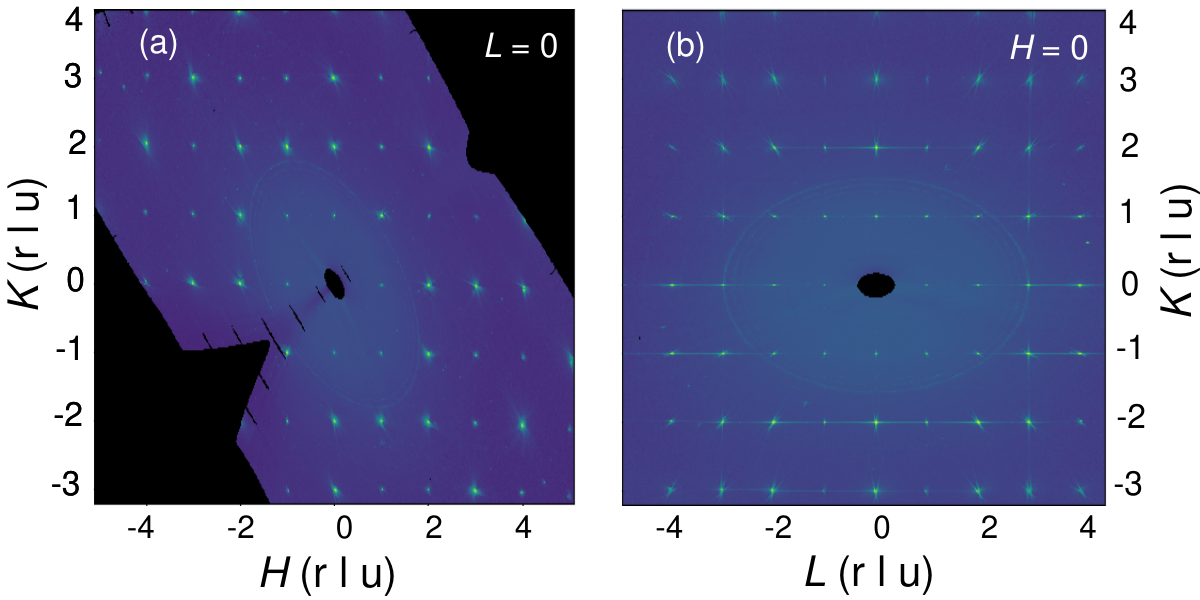}
	\caption{Synchrotron X-ray diffraction data. (a) Bragg reflections in the (H, K, 0) scattering plane. (b) Bragg reflections in the (0, K, L) scattering plane. Axis units are plotted in r.l.u.
	}
	\label{Fig_1_1}
\end{figure}

\section{Results}

\subsection{X-ray diffraction and crystal structure}

The crystal structure of TbV$_6$Sn$_6$, obtained from the 300 K refinement of single crystal X-ray diffraction data in space group $P6/mmm$ is shown in Fig. \ref{Fig_1}. The room temperature  model (via a laboratory X-ray source) agrees with the refinement of 300 K synchrotron data, and low-temperature synchrotron data show no symmetry breaking down to 10 K. Tb and V-ions occupy the unique hexagonal crystallographic sites 1b and 6i respectively; whereas Sn occupy three different hexagonal crystallographic sites 2d(Sn1), 2e(Sn2) and 2c(Sn3). The structure consists of layered units of [V$_3$Sn1][TbSn2][V$_3$Sn1][Sn3] stacking along the crystallographic c-axis. Fig \ref{Fig_1}(b) presents a topside view of the crystal structure, and it highlights the kagome layer of V-atoms within the ab-plane. Each unit cell is made by a slab of V-Sn1-Sn3-Sn1-V atoms and a TbSn2 layer. In the TbSn2 layer, Tb atoms occupy the center of a stanene lattice of Sn2 atoms. The interstitial Tb atoms form a triangular lattice network as shown in Fig. \ref{Fig_1}(c).

The structural parameters at 300 K, 100 K, and 10 K are shown in Table \ref{table}. Due to the presence of interlayer diffuse scattering in the synchrotron X-ray data, the thermal parameters of V-sites exhibited unstable behavior while optimizing a structural model during refinement. To compensate, the thermal parameters of V-sites determined via laboratory X-ray diffraction refinement at room temperature were used and subsequently fixed while refining the synchrotron data. At 300 K, the refined nearest-neighbor (NN) V-V distance of 2.777(2) \AA {} within a kagome plane is much smaller than the out-of-plane V-V distance of 4.633(3) \AA {} along the c-axis. Magnetic ions of Tb occupy the corners of the unit cell, yielding a distance between NN Tb ions equal to the value of lattice parameter in all directions. Upon cooling, the unit cell, V-V, Tb-Tb distances all contract in a conventional fashion.

 Figs. \ref{Fig_1_1}(a) and (b) display 300 K synchrotron X-ray scattering data highlighting select Bragg peaks in the ($H$, $K$, 0) and (0, $K$, $L$) scattering planes, respectively. Diffuse scattering is apparent along the $L$ direction, likely arising from interlayer structural disorder. Correlation lengths within the plane and between the planes do not change appreciably upon cooling to 10 K. Representative minimum correlation lengths $\xi_{min}$ were extracted from the $(-1,0,4)$ peak by fitting resolution-limited Gaussian lineshapes along the $H$ and $K$ directions. These fits render $\xi_{H/K}=900(20)$~\AA in the ab-plane.  Deconvolving the resolution via fits to a Voigt lineshape for cuts along the $L$-axis.  These yielded an interplanar correlation length $\xi_{L}=460(10)$ ~\AA.

\begin{table}[h!]
\caption{Structural details of TbV$_6$Sn$_6$ obtained from the refinement of single crystal X-ray diffraction data in space group {\it $P6/mmm$}. 
}

	\begin{tabular}{ccccc}
		& & T = 300 K  & & 
		\end{tabular}
    \renewcommand{\tabcolsep}{5.5pt}
	\begin{tabular}{ccccc}
		\hline 
		$a$ = $b$ (\AA) & $c$ (\AA)  & $V$ (\AA$^3$)&  $c/a$   \\ \hline
		5.5346(3) & 9.2065(7) & 244.23(3) & 1.6634  \\
		\end{tabular}
	\renewcommand{\tabcolsep}{1.5pt}
	\begin{tabular}{c|cccccc}
		\hline 
	     Atom & site & $x$ & $y$ & $z$ & $U_{iso} (\AA^2)$  & $Occupancy$ \\ \hline
		Tb & 1b & 0 & 0 & 0           & 0.0072(10)   & 1 \\ \hline
		Sn1 & 2d  & 0 & 0 & 0.33364(17)             & 0.0070(11)  & 1 \\ \hline
		Sn2 & 2e  & 0.333333 & 0.666667 & 0           & 0.0047(10)   & 1 \\ \hline
		Sn3 & 2c  & 0.666667 & 0.333333 & 0.500000             & 0.0057(10)  & 1 \\ \hline
		V & 6i  & 0.500000 & 0.500000 & 0.24838(16)    & 0.007   & 1 \\ \hline 
		\end{tabular}
		
		\hskip+0.3cm
		
	\begin{tabular}{ccccc}
		& & T = 100 K & & 
		\end{tabular}
    \renewcommand{\tabcolsep}{5.5pt}
	\begin{tabular}{ccccc}
		\hline 
		$a$ (\AA) & $c$ (\AA)  & $V$ (\AA$^3$)&  $c/a$  \\ \hline
		5.5180(3) & 9.1746(9) & 241.93(4) & 1.6626 \\
		\end{tabular}
	\renewcommand{\tabcolsep}{1.5pt}
	\begin{tabular}{c|cccccc}
		\hline 
		Atom & site & $x$ & $y$ & $z$ & $U_{iso} (\AA^2)$ & $Occupancy$ \\ \hline
		Tb & 1b & 0 & 0 & 0           & 0.0028(11)   & 1 \\ \hline
		Sn1 & 2d & 0 & 0 & 0.33409(19)             & 0.0077(14)  & 1 \\ \hline
		Sn2 & 2e & 0.333333 & 0.666667 & 0           & 0.0044(16)   & 1 \\ \hline
		Sn3 & 2c & 0.666667 & 0.333333 & 0.500000             & 0.0044(16)  & 1 \\ \hline
		V  & 6i & 0.500000 & 0.500000 & 0.24823(19)    & 0.007   & 1 \\ \hline 
		\end{tabular}

	\hskip+0.3cm

	\begin{tabular}{ccccc}
		& & T = 10 K & & 
		\end{tabular}
    \renewcommand{\tabcolsep}{5.5pt}
	\begin{tabular}{ccccc}
		\hline 
		$a$ (\AA) & $c$ (\AA)  & $V$ (\AA$^3$)&  $c/a$  \\ \hline
		5.5124(3) & 9.1655(9) & 241.19(4) & 1.6627 \\
		\end{tabular}
	\renewcommand{\tabcolsep}{1.5pt}
	\begin{tabular}{c|cccccc}
		\hline 
		Atom & 	site & $x$ & $y$ & $z$ & $U_{iso} (\AA^2)$ & $Occupancy$ \\ \hline
		Tb & 1b  & 0 & 0 & 0           & 0.0013(14)   & 1 \\ \hline
		Sn1 & 2d & 0 & 0 & 0.3343(3)             & 0.0055(17)  & 1 \\ \hline
		Sn2 & 2e & 0.333333 & 0.666667 & 0           & 0.0001(18)   & 1 \\ \hline
		Sn3 & 2c & 0.666667 & 0.333333 & 0.500000             & 0.0002(18)  & 1 \\ \hline
		V & 6i   & 0.500000 & 0.500000 & 0.2483(3)    & 0.007   & 1 \\ \hline \hline
		\end{tabular}
		\label{table}

	\end{table}

\subsection{Magnetic properties}

\begin{figure} 
\centering
\includegraphics[width=1\columnwidth]{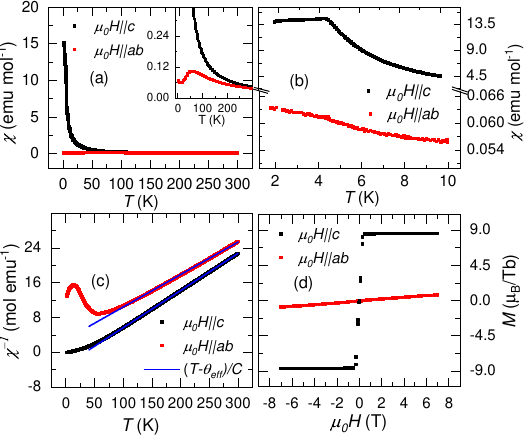}
      \caption{Magnetic properties of TbV$_6$Sn$_6$. Red and black squares represent the data measured with the applied field parallel and perpendicular to the (001) crystal surface. (a) Temperature dependent magnetic susceptibility, $\chi$, at an applied field of 100 mT. The inset of (a) displays a broad anomaly around 60 K in $\chi$ along the (001) crystal surface. (b) Magnetic susceptibility below 10 K measured with a field of 10 mT. (c) Inverse magnetic susceptibility at an applied magnetic field of 100 mT. (d) Field dependent isothermal magnetization at 1.8 K.
      }
\label{Fig2}
\end{figure}

Magnetization data collected as a function of temperature and magnetic field are summarized in Fig. \ref{Fig2}. For each set of data, the magnetic field is applied either parallel or perpendicular to the $(0, 0, 1)$ axis, and the magnetization exhibits a strongly anisotropic behavior. At low temperature the magnetic susceptibility, $\chi$, along the $c$-axis is approximately two orders of magnitude weaker than its value within the ab-plane (Fig. \ref{Fig2}(b)). Just above the ordering temperature at $T=5$ K, the anisotropy reaches a value of $\chi_{c}/\chi_{ab}=197$ under a 10 mT field. This demonstrates a strong easy-axis anisotropy in TbV$_6$Sn$_6$, similar to a number of other Tb-based compounds \cite{PhysRevB.53.307,PhysRevB.95.134434,Chern_Yin2020_TbV6Sn6,Frontzek2006MagnetocrystallineAI}. The temperature-dependent magnetization reveals a magnetic transition at $T_C$ = 4.3(2) K, shown most clearly in the c-axis aligned data in Fig. \ref{Fig2}(b). 

Isothermal, field-dependent magnetization data collected below $T_C$ are plotted in Fig. \ref{Fig2}(d). Along the easy-axis direction, the moment polarizes rapidly and reaches saturation near 1 T. The value of the saturated moment along the easy axis is nearly 9 $\mu_B$, as expected for the $4f^8$ electronic configuration of Tb$^{3+}$ ions with $J=6$ and $g_{Lande}=1.5$. When the field is aligned along the ab-plane, the moment only weakly responds and fails to saturate up to 7 T. This demonstrates that TbV$_6$Sn$_6$ is an extreme easy-axis ferromagnet in which the value of g-tensor is nearly zero perpendicular to the moment easy-axis. The ferromagnetic state is rather soft, and the coercivity is $\approx0.015$ T.       

Apart from $T_C$, an additional broad magnetic anomaly appears in the temperature-dependent magnetization data near 60 K. The inset of Fig. \ref{Fig2}(a) illustrates this broad feature in $\chi$, which appears only when the magnetic field is oriented within the ab-plane. This broad feature is consistent with the presence of a low-lying crystal field of the split $J=6$ ground state multiplet.  Upon heating, thermal occupation of the first excited state drives an excitonic mixture of a multiplet with greater easy-plane character, and, above this temperature, the susceptibility becomes increasingly isotropic.   

The inverse magnetic susceptibility, $\chi^{-1}$, is plotted in Fig. \ref{Fig2}(c). $\chi^{-1}$ varies linearly with temperature above 100 K once the first low-lying crystal field multiplet is appreciably populated. An empirical Curie-Weiss analysis, $\chi^{-1} =  (T-\Theta_{eff})/C$; $C$ = Curie constant, $\Theta_{eff}$ = an empirical Curie-Weiss temperature in the high temperature, thermally mixed state, was performed above 100 K and the fits are overplotted with the data in Fig. \ref{Fig2}(c). This analysis yields $\Theta_{eff}=33.1(1)$ K parallel and $\Theta_{eff}=-40.4(2)$ K perpendicular to the moment easy-axis in the excitonic regime.  Curie constants of 11.69(1) and 13.41(1) cm$^3$ mol$^{-1}$ are fit with the field parallel and perpendicular to the c-axis respectively. These Curie constants translate to effective magnetic moments of 9.6 and 10.3 $\mu_B$/Tb along the respective directions, consistent with the $J$= 6 ($L$=3, $S$=3) moment of Tb$^{3+}$ ions. Higher magnetic fields or a solution for the non-Kramers ground state multiplet wave functions will be required to quantitatively determine the precise g-tensor anisotropy.

\subsection{Transport properties}
\label{Trp}

\begin{figure} 
\centering
\includegraphics[width=1\columnwidth]{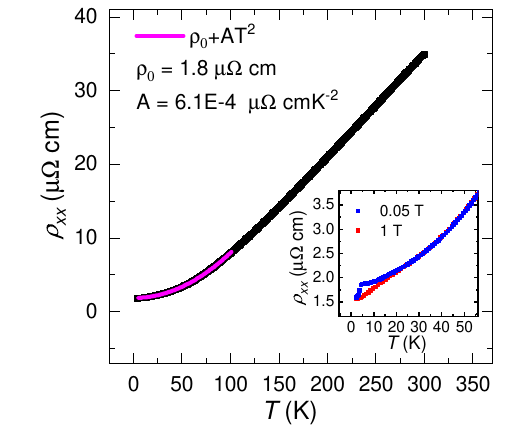}
      \caption{Low temperature electrical resistivity $\rho_{xx}(T)$ of TbV$_6$Sn$_6$ around $T_C$ at fields of 0.05 and 1 T applied along the crystallographic $c$-axis. Inset shows $\rho_{xx}(T)$, in the temperature range of 4.5 - 300 K, measured with a current flowing in the $ab$-plane at zero magnetic field. The pink color curve in the inset shows a fit of data to the equation $\rho (T) = \rho(0) +AT^2$           . 
      }
\label{Fig_3}
\end{figure}

Temperature dependent electrical resistivity $\rho_{xx} (T)$ data measured under magnetic fields of $\mu_0 H$ = 0, 0.05 and 1 T are displayed in Fig. \ref{Fig_3}.  $\rho_{xx}$ decreases smoothly upon cooling toward $T_C$, demonstrating a residual resistivity ratio $\rho_{xx}(300$ K$)/\rho_{xx}(2$ K$)\approx18$. Zero-field data show a weak decrease near 3.5 K associated with Sn-superconductivity.  This effect arises from small quantities of Sn-flux inclusions likely within the bulk of the crystal. This effect is removed by applying a field larger than the critical field ($\approx 300$ Oe) of Sn, and after doing so, TbV$_6$Sn$_6$ exhibits an inflection in the value of $\rho_{xx}(T)$ when cooling below $T_C$. The inset of Fig. \ref{Fig_3} illustrates this small drop in resistivity at $T_C$, and when the ordered state is fully saturated ($\mu_0 H$= 1 T), the inflection about $T_C$ becomes significantly broader. 

The resistivity data in the paramagnetic regime below 100 K is well fit to the form $\rho_{xx} (T)$ = $\rho(0) + AT^2$.  The fit is overplotted with the data in Fig. \ref{Fig_3} and gives $\rho_{xx}(0)=1.884(3)$ $\mu\Omega$ cm and $A$ = 6.154(5)$\times$ 10$^{-4}$ $\mu\Omega$ cm K$^{-2}$. The values of $\rho_{xx}(0)$ and $A$ are close to the values of 2 $\mu\Omega$ and 3.5 $\times$ 10$^{-4}$ $\mu\Omega$ cm K$^{-2}$ for similar materials GdV$_6$Sn$_6$ and YV$_6$Sn$_6$ respectively \cite{Pokharel_2021,Japanese_166}. At high temperatures near 300 K, $\rho_{xx}(T)$ switches to follow a conventional, linear temperature dependence.

\begin{figure} 
\centering
\includegraphics[width=1\columnwidth]{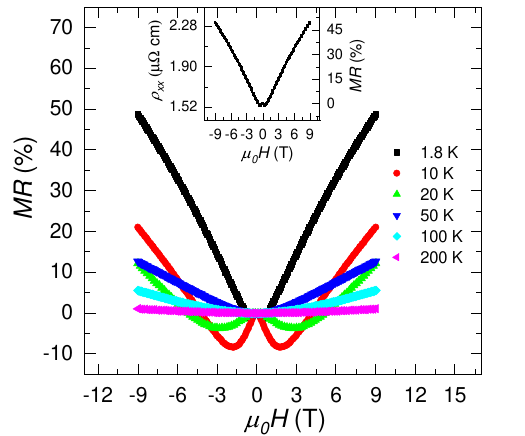}
      \caption{Transverse magnetoresistance data. Isothermal magnetoresistance ratio (MR) increases with the decrease in temperature and it exhibits a minima at low temperature due to the coupling with the  spin polarization. The inset shows the magnetoresistance data at 1.8 K. The isothermal $\rho_{xx}$(H) data exhibited weak asymmetry upon reversed field direction due to the contamination from Hall signal. The magnetoresistance component was isolated by averaging $\rho_{xx}(H)$ and $\rho_{xx} (-H)$. At high fields, $\rho_{xx}$ displays linear field dependence.    
      }
\label{Fig_4}
\end{figure}

The field dependence of the transverse magnetoresistance $\rho_{xx}(H)$ collected at various temperatures and the resulting magnetoresistance ratio (MR) are plotted in Fig. \ref{Fig_4}.  $\rho_{xx}(H)$ reveals a weak asymmetry with the reversal of the applied field direction due to the mixing of the transverse magnetoresistance with a strong Hall signal. The magnetoresistance is isolated from the mixed Hall signal by extracting the symmetric component of $\rho_{xx}(H)$ via averaging over the negative and positive field directions \cite{Feng_2015, Visser_2006}. The field dependence of resistance changes from negative and quadratic at low fields to linear and positive at high fields, and the resulting minima in the MR at low temperature form as the field polarizes the Tb moments and suppresses spin fluctuations.  At 1.8 K, positive MR appears near 1 T, the field where TbV$_6$Sn$_6$ is driven into a fully spin polarized state. Nonsaturating, positive MR is observed up to the maximum applied field of 9 T, where the MR reaches $\approx 50$ \% at 1.8 K. .  

\begin{figure}[h]
\centering
\includegraphics[width=1\columnwidth]{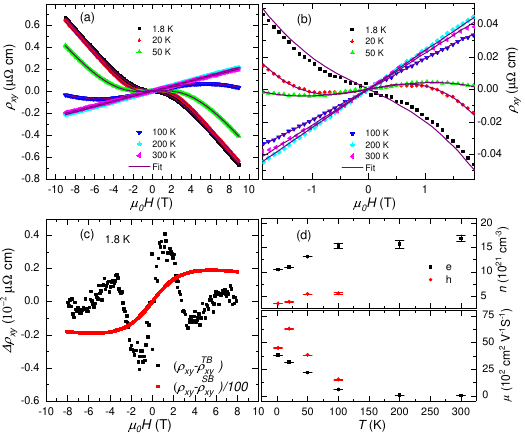}
      \caption{Hall effect measurements on TbV$_6$Sn$_6$. (a) Hall resistivity,$\rho_{xy}$, as a function of magnetic field, measured at different temperatures. To remove mixing of transverse resistivity, $\rho_{xy}$ is isolated as $\rho_{xy} =  (\rho_{xy}(H)-\rho_{xy}(-H)$)/2. Fits to a two-band model below 100 K and single band model above 200 K are represented by solid lines through the data. (b) $\rho_{xy}$ at low field at various temperatures focusing on the deviation of model at 1.8 K. (c) The anomalous hall signal and non-linear Hall resistivity, $\Delta \rho_{xy}$, at 1.8 K. The calculated resistivity using two-band (TB) model and single-band (SB) model based on the linear region at high field is subtracted to calculate $\Delta \rho_{xy}$. (d) Upper panel shows the temperature dependence of electron-type and hole-type carrier densities, $n$, obtained by the fits to the Hall data. e and h in the figure represents electron- and hole-type charge carriers. Lower panel of (d) shows the temperature dependence of the charge carrier mobilities, $\mu$ for each carrier type obtained from fits to the Hall data.  
      }
\label{Fig_5}
\end{figure}

Hall resistance $\rho_{xy}(T,H)$ data were collected to characterize the carrier concentrations and effective mobilities of the conduction bands. Fig. \ref{Fig_5} shows $\rho_{xy}(H)$ measured at various temperatures with the magnetic field applied parallel to the c-axis. At 200 K and above, $\rho_{xy}(H)$ shows a linear field dependence, indicating a dominant single-carrier conduction channel. As the temperature is lowered, $\rho_{xy}$ switches to a nonlinear form, consistent with a multi-carrier model.  As a result, $\rho_{xy}(H)$ was fit with two different models: a two-band model at low ($T < 100$ K) and a single band model at high ($T > 200$ K).  
 
%
%
%

 Fits to $\rho_{xy}$ are shown by solid lines in Fig. \ref{Fig_5}(a). With the exception of the low-field data at 1.8 K (below $T_C$), the fits capture the data well. The deviation from the two-band model below $T_C$ is visible in Fig. \ref{Fig_5}(b), and likely represents the onset of the anomalous Hall component within $\rho_{xy}$. Attempting to extract the anomalous Hall signal $\Delta\rho_{xy}$ via subtracting the two-band model from the experimental data is plotted in Fig \ref{Fig_5}(c); however it instead reveals an oscillatory component within the Hall signal.  This could correspond to Hall data reflecting quantum oscillations arising from a small Fermi pocket, though the oscillations are not readily resolved in magnetoresistance $\rho_{xx}$ data \cite{kikugawa2010ca3ru2o7}.  An alternative analysis is to instead use a single band treatment and propagate the 1.8 K, high-field Hall response backward linearly and extract an anomalous Hall signal from the difference.  This is also plotted in Fig \ref{Fig_5}(c), and, while the difference provides a signal more consistent with a traditional anomalous Hall effect, it also neglects the multiband character of the low-$T$ transport. A similar analysis using this single band treatment with nonmagnetic YV$_6$Sn$_6$, for instance, resolves a qualitatively similar signal \cite{Japanese_166}. Future, detailed electronic structure calculations and lower temperature measurements will be important for determining the intrinsic magnitude of the anomalous hall effect (AHE) in TbV$_6$Sn$_6$. 
 
 Above $T_C$ and below 100 K, two-band fits capture $\rho_{xy}$ well and both electron-like and hole-like carriers contribute to the Hall signal. The resulting electron-type carrier density, $n_e$, and hole-type carrier density, $n_h$, are plotted in the upper panel of Fig. \ref{Fig_5}(d). The mobilities of each are plotted within the lower panel of Fig. \ref{Fig_5}(d), where each carrier type's mobility becomes comparable at the lowest measured temperature.  Above 200 K, $\rho_{xy}$ data are dominated by a single carrier type and a single band model showing electron-like carriers whose mobility increases quickly upon cooling. 

\subsection{Heat capacity measurements}

\begin{figure} 
\centering
\includegraphics[width=1\columnwidth]{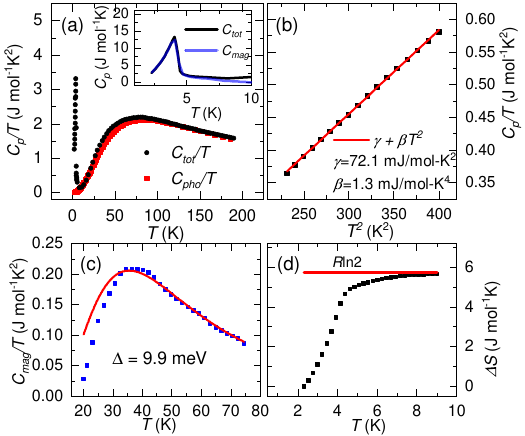}
      \caption{Heat capacity, $C_p(T)$, measurements in TbV$_6$Sn$_6$. (a) Temperature dependence of total ($C_{tot}/T$) and phonon ($C_{pho}/T$) heat capacity observed at zero-magnetic field. The inset of (a) represents total ($C_{tot}$) and magnetic ($C_{mag}$)  heat capacity around $T_C$. $C_p(T)$ displays a sharp $\lambda$-like anomaly at 4.1 K. (b) Temperature dependence of $C_p/T$ plotted as a function of $T^2$. Solid red color line shows the fit to the Sommerfeld form  $C_p(T)=\gamma T$ + $\beta T^3$ in paramagnetic region above $T_C$. (c) Magnetic contribution to the heat capacity, $C_{mag}/T$, in the temperature range of 20-75 K.  Solid line is a fit to a two-level Schottky anomaly as described in the text. (d) Magnetic entropy change, $\Delta S$, at low temperature below 9 K. The red color line represents the calculated entropy for an effective spin 1/2 system.    
      }
\label{Fig6}
\end{figure}

Temperature-dependent heat capacity data, $C_{p}(T)$, collected between 1.8 K and 300 K at zero field are plotted in Fig.~\ref{Fig6}. $C_{p}(T)$ data exhibit a sharp anomaly near 4.1~K, coinciding with the onset of magnetic order below $T_{C}$. The low-temperature $C_{p}(T)$ collected away from short-range magnetic fluctuations obeys the Sommerfeld form $C_{p}$ = $\gamma T$ + $\beta T^{3}$, and the resulting fit is plotted in Fig. \ref{Fig6}(b). The fit yields the coefficients $\gamma=72.1(20)$~mJ~mol$^{-1}$ K$^{-2}$ and $\beta=1.28(1)$~mJ~mol$^{-1}$~K$^{-4}$ translating to a $\Theta_{D}\approx111$ K---though the $\gamma$ value extracted is purely phenomenological.  The magnetic contribution to the heat capacity, $C\rm{_{mag}}$, through the ordering temperature can be isolated by subtracting the approximate phonon contribution, derived from measurements of YV$_6$Sn$_6$. This is overplot with TbV$_6$Sn$_6$ data in the inset of Fig. \ref{Fig6}(a). Fig. 7 (d) shows the integration of the magnetic entropy $\Delta S$ determined from this data from 2 K to 9 K.  The value approaches that expected for a non-Kramers doublet ground state with $S=Rln(2)$; however there remains unaccounted magnetic entropy below 2 K that is not included in this sum.    

$C\rm{_{mag}}$ at higher temperatures is plotted in Fig. \ref{Fig6}(c). A broad hump appears near 40 K, consistent with a Schottky anomaly expected from the thermal population of a low lying crystal field state. A little discrepancy to the energy scale of first exited crystal field state, compared to the energy scale observed in magnetization measurements, may be associated with the large size difference of the R-site ions and the systematic errors associated with the measurements. We believe that the subtraction of phonon contributions to the heat capacity is not perfect due to the large size difference of Y and Tb-ions and stronger contribution of Einstein heat capacity in YV$_6$Sn$_6$. A simple, two level model is used to fit the data to the form $C_{Schottky} = s\times R(\frac{\Delta}{T})^2\frac{e^{\Delta/T}}{[1+e^{\Delta/T}]^2}$ with $s$ being an overall scale factor and $\Delta$ being the energy gap between the two-level system.  This form captures the asymmetric peak reasonably well on the high temperature side; however, upon cooling $C\rm{_{mag}}$ drops more rapidly than expected.  This misfit at the low temperature portion of the Schottky peak likely arises from a slight oversubtraction of the phonon contribution using the scaled YV$_6$Sn$_6$ data.  The parameterization predicts the first excited state near $\Delta=114$ K, though determining the full level scheme will require either inelastic neutron or Raman spectroscopy.  

\section{Discussion}
\label{discussion}

The strong easy-axis anisotropy observed in TbV$_6$Sn$_6$ is reasonably common for Tb ions in a strong crystalline electric-field (CEF) \cite{PhysRevB.95.134434,doi:10.1143/JPSJS.80SA.SA080, doi:10.1080/14786430903074797}. The relaxation of the magnetic anisotropy upon warming suggests that significant thermal population of the first $J=6$ intramultiplet excitation occurs beyond 60 K, and the first excited state is identified at $\Delta=114$ K via $C_p$ data. Inelastic neutron scattering measurements on phase-pure powders or scaled volume single crystals are the natural next step for fully determining the $J=6$ intramultiplet excitation scheme.  Neutron diffraction measurements may also provide useful for fully parameterizing the magnetic ground state (e.g. ordered moment, zero field orientation, ...).
 
First principles calculations of the electronic band structure of RV$_6$Sn$_6$ show a dominant V $3d$ contribution at the Fermi level \cite{Japanese_166, Pokharel_2021}, indicating the transport properties of TbV$_6$Sn$_6$ are dominated by the V-kagome layer. In the paramagnetic state, metallic state can be characterized by examining the Kadowaki-Woods ratio ($A/\gamma^2$), and the ratio $A/\gamma^2=8.5\times10^{-3}$ $\mu\Omega$ cm mol$^2$ K$^2$ J$^{-2}$ for TbV$_6$Sn$_6$ is lower than the values for strongly correlated heavy fermion systems, suggesting only moderate correlation effects \cite{KADOWAKI_1986, Jacko2009}.  

A weak, low-field departure from a conventional two-band model below $T_C$ likely corresponds to the onset of the anomalous Hall effect in TbV$_6$Sn$_6$. It is however difficult to reliably isolate the contribution given the small hysteresis of the ferromagnetic state, small temperature range accessible in the ordered state, and the added complication of a two-band transport background.  The absence of a strong mean exchange field created by magnetic $3d$ transition metal ions on the kagome lattice softens the ordered state, and future studies isolating the anomalous Hall component will provide an interesting comparison to the quantum-limit Chern magnetism reported in TbMn$_6$Sn$_6$ \cite{Chern_Yin2020_TbV6Sn6, Xu2022}. Furthermore, by isolating the magnetic order within Tb planes, TbV$_6$Sn$_6$ affords independent control of the anisotropy and the character of magnetic order via alloying the $R$-site while retaining the electronic structure of the nonmagnetic V-kagome layers.

\section{Conclusions} 
The synthesis and characterization of single crystals of the kagome metal TbV$_6$Sn$_6$ were presented, and a c-axis aligned, easy-axis ferromagnetic state is reported to form below $T_C=$4.3(2) K. Magnetotransport measurements reveal high-mobility, multiband transport, with a departure observed below $T_C$ likely signifying the onset of an anomalous Hall component. Our work demonstrates that $R$V$_6$Sn$_6$ compounds allow for tunable anisotropies and magnetic states on the $R$ sites of $R$V$_6$Sn$_6$ kagome metals while preserving a nonmagnetic kagome lattice. In the present case, strong uniaxial, out-of-plane ferromagnetism can be engineered via the selection of the Tb ion. In contrast to the isotropic Gd ions or in-plane magnetism of the Er and Tm ions in the R sites \cite{02924, Zhang}, TbV$_6$Sn$_6$ exhibits strong uniaxial out-of-plane magnetism that saturates quickly at low magnetic field below 1 T. 

\section{Note added} 

Our results are largely consistent with recent works that became available while this manuscript was in review \cite{14802,02924}. In Rosenberg et al. \cite{14802}, strong uniaxial ferromagnetism was observed as well as multiband transport consistent with our findings.  A strong gap between the ground state doublet and first excited crystal field state was postulated, consistent with our $C_p$ data.  In Lee et al. \cite{02924} similar heat capacity data were reported, although analysis of the magnetic TbV$_6$Sn$_6$ transition in that work reports a zero field antiferromagnetic transition.  This is based on powder averaged, high temperature Curie-Weiss fits to the data that extract a antiferromagnetic exchange field.  Our analysis views this high temperature regime as one with mixed crystal field multiplets that are thermally occupied, and, that the low temperature, non-Kramers doublet ground state realizes ferromagnetic order. In Zhang et al. \cite{Zhang} similar multiband transport, heat capacity data, and strong uniaxial ferromagnetism was reported.

\begin{acknowledgments}
This work was supported via the UC Santa Barbara NSF Quantum Foundry funded via the Q-AMASE-i program under award DMR-1906325. We acknowledge the use of the computing facilities of the Center for Scientific Computing at UC Santa Barbara supported by NSF CNS 1725797 and NSF DMR 1720256. BRO and PMS acknowledge financial support from the UC, Santa Barbara through the Elings Fellowship. This work is based upon research conducted at the Center for High Energy X-ray Sciences (CHEXS) which is supported by the National Science Foundation under award DMR-1829070.  Any opinions, findings, and conclusions or recommendations expressed in this material are those of the authors and do not necessarily reflect the views of the National Science Foundation. 
\end{acknowledgments}


%

\end{document}